\documentclass[10pt,twocolumn]{article}
\usepackage[a4paper]{geometry}
\usepackage{mathptmx}

\usepackage{listingsutf8}
\usepackage[utf8]{inputenc}
\usepackage[T1]{fontenc}

\usepackage{amssymb,amsmath}
\usepackage{graphicx}
\usepackage{blkarray}
\usepackage{dsfont}
\usepackage{multirow}
\usepackage[table]{xcolor}
\usepackage{subfig}
\usepackage{relsize}
\usepackage{sidecap, caption}
\usepackage{comment}
\usepackage[bottom]{footmisc}
\newcommand{\boldm}[2] {\mathversion{bold}#1\mathversion{normal}}
\usepackage{upgreek}
\usepackage[hyphens]{url}
\usepackage[colorlinks=true,linkcolor=blue,
	citecolor=blue,urlcolor=blue]{hyperref}

\usepackage{color,xcolor}
\definecolor{metal}{RGB}{63,63,63}
\definecolor{rouge}{RGB}{220,20,60}
\definecolor{vert}{rgb}{0.02,0.53,0}
\definecolor{bleu}{RGB}{0,0,139}

\usepackage{ntheorem}

\makeatletter
\renewtheoremstyle{plain}
  {\item{\theorem@headerfont ##1\ ##2\theorem@separator}~}
  {\item{\theorem@headerfont ##1\ ##2\ (##3)\theorem@separator}~}
\makeatother

{\theoremheaderfont{\upshape\bfseries}
 \theorembodyfont{\normalfont\slshape}
 \newtheorem{definition}{Definition}}

\newtheorem{problem}{Problem}

\newtheorem{remark}{Remark}

\renewcommand{\thesection}{\Roman{section}}
\renewcommand{\thesubsection}{\Alph{subsection}}
\renewcommand{\thesubsubsection}{\arabic{subsubsection}}
\makeatletter
\renewcommand*{\p@subsection}{\thesection\hspace{0.02cm}}
\makeatother

\usepackage{titlesec}
\titleformat{\section}[hang]{\centering\bfseries}{\thesection.}{0.5em}{}
\titleformat{\subsection}[hang]{\centering\bfseries}{\thesubsection.}{0.5em}{}
\titleformat{\subsubsection}[hang]{\centering\itshape}{\thesubsubsection.}{0.5em}{}

\usepackage{authblk}
\usepackage{blindtext}
\title{\textbf{Link weights recovery in heterogeneous information networks}}
\author[1]{Hong-Lan Botterman}
\author[2]{Robin Lamarche-Perrin}
\affil[1]{{\small \textit{Sorbonne Universit\'{e}, CNRS, LIP6, F-75005 Paris, France}} \vspace{-0.1cm}}
\affil[2]{{\small \textit{Institut des syst\`{e}mes complexes de Paris \^{I}le-de-France, ISC-PIF, UPS 3611, Paris, France}}}
\date{}

\usepackage{xcoffins,titling}
\NewCoffin \myabstractcoffin
\newcommand \myabstract[2][.8]{%
  \SetVerticalCoffin \myabstractcoffin {#1\textwidth} {#2}%
  \renewcommand\maketitlehookd{%
    \centering
    \TypesetCoffin \myabstractcoffin}}
    
\usepackage{fancyhdr}
\begin{document}

\myabstract{
Socio-technical systems usually consists of many intertwined networks, 
each connecting different types of objects (or actors) through a variety 
of means. As these networks are co-dependent, one can take advantage of 
this entangled structure to study interaction patterns in a particular 
network from the information provided by other related networks.
A method is hence proposed and tested to recover the weights of missing or unobserved links in heterogeneous information networks (HIN) - abstract representations of systems composed of multiple types of entities and their relations.
Given a pair of nodes in a HIN, this work aims at recovering the exact weight of the incident link to these two nodes, knowing some other links present in the HIN. 
To do so, probability distributions resulting from path-constrained random walks i.e., random walks where the walker is forced to follow only a specific sequence of node types and edge types, capable to capture specific semantics and commonly called a meta-path, are combined in a linearly fashion in order to approximate the desired result.  
This method is general enough to compute the link weight between any types of nodes. Experiments on Twitter and bibliographic data show the applicability of the method.
}
\maketitle
\thispagestyle{empty}

\section{INTRODUCTION}
\label{intro}
\noindent Networked data are ubiquitous in real-world applications. Examples of such data are humans in social activities, proteins in biochemical interactions, pages of Wikipedia or movies-users from Amazon just to name a few. These are abstracted by a network where nodes represent the entities (e.g. individuals or pages) of the examined system whilst (directed) links stand for existing physical or virtual ties between them. Weights can also be put on links to state, for instance, their importance.
In some cases, the nodes and/or the links are of different nature. For example, in social activities, the links can reflect online or offline communication or more obviously, in the movie-user case, nodes represent two different objects. 
Taking these differences explicitly into account in the modeling can only enrich the understanding of the inspected system. 
Thus, heterogeneous information networks (HIN), abstract representations of systems composed of multiple types of entities and their relations, are good candidates to model such data together with their relations, since they can effectively fuse a huge quantity of information and contain rich semantics in nodes and links.

In the last decade, the heterogeneous information network analysis has attracted a growing interest and many novel data mining tasks have been designed in such networks, such as similarity search, clustering, classification and link prediction \cite{shi2016survey}. 
The latter can sometimes refer to the term recovery, in the sense that links already exist but are missing or imperfectly  observed in the data. This could be due to sampling or depending on the system under scrutiny, due to node/agent's voluntary decision not to give access to all her data (e.g. online social apps).  
Whatever the reason, capturing the presence of a link is sometimes not enough sufficient. For instance, in a social network, knowing two individuals are linked does not say anything about the frequency of their communication or the strength of their friendship. 
Hence recovering the actual link weight can bring useful information as for instance, in recommendation systems where the weight can be taken for the ``rating" a user would give to an item. 
The goal of this work is to recover, for a given pair of nodes in a weighted HIN, the actual incident link weight to these two nodes, knowing some other links present in the HIN. 

Link prediction can be related to node similarity problem. Indeed, the similarity score between two nodes, result of a particular function of these two nodes, can be seen as the strength of their connection.
Here, this function is related to a particular random walk on the graph and so, to the probabilities of reaching one node through different paths, starting from another.

In HIN, most of similarity scores \cite{lao2010relational,sun2011co} are based on the concept of meta-path. In simple terms, this corresponds to a concatenation of node types linked by corresponding link types and the type of a node/link is basically a label in the abstract representation. 
Meta-paths can be used as a constraint to a classic random walk: the walker is allowed to take only paths satisfying a particular meta-path. These path-constrained random walks have the sensitivity to explicitly take into account different semantics present in HIN.
For instance, in a bibliographic network, we can distinguish four types of entities: Authors (A), Papers (P), Venues (V) and Topics (T). Starting from a particular paper, if a walker follows the meta-path PVP, he is likely to end to any another paper published in the same venue than the first. Now, if he follows the meta-path PTP, the ending paper will be about the same topic. Even if the starting and ending papers are the same, the semantics behind may be radically different.

Back to our goal, we can see it as a (linear) regression problem where the aim is to recover the link weight i.e., a continuous value.
This means that the target link weight between a pair of nodes is approximated by a linear combination of probabilities, results of path-constrained random walks performed on the HIN. These probabilities thus translate the fact of being at a particular node starting from another one and are the regressors  for the linear regression.   
Thenceforth, in order to make recovery tasks, data is commonly split into two sets: training and test. The proposed method aims at finding a relevant set of meta-paths together with their coefficient such that the difference between the exact link weight and its approximation is minimized for the training set. Obtained coefficient are then tested on the test set. 
 
The rest of this paper is organized as follows. 
In Sec. \ref{Concepts}, some basic concepts about HIN are presented and the problem statement is exposed.
Sec. \ref{Method} explains our method and we apply it on empirical data in Sec. \ref{Experiments}. First, in Sec. \ref{Experiments_Different}, the method is tested to recover the link weights between entities of different types into Twitter data. Then, in Sec. \ref{Experiments_Same}, it is applied on bibliographic data with the same type of target nodes. 
We review some related work in Sec. \ref{RelWork} and  we finally conclude and discuss some perspectives in Sec. \ref{Conclusion}.

\section{PREMIMINARY CONCEPTS}
\label{Concepts}

\begin{figure*}[h]
\captionsetup[subfigure]{position=top, labelfont=bf,textfont=normalfont,singlelinecheck=off,justification=raggedright}
\centering
\begin{tabular}{cccc}
\subfloat[]{\includegraphics[scale=1]{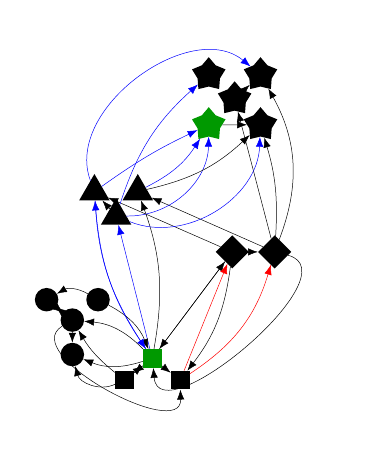} \label{IllDefa}}
&
\subfloat[]{\includegraphics[scale=1]{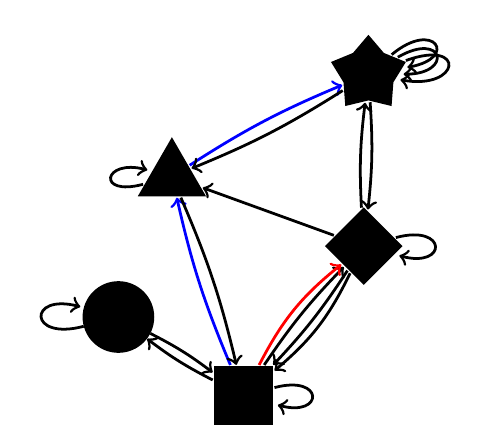} \label{IllDefb}}
&
\hspace{0.2cm}
&
\subfloat[]{
\vspace*{2cm}
\begin{tabular}{l}
$\textcolor{red}{E_c} = \blacksquare \rightarrow \blacklozenge$  \\ \\
\hspace{-0.2cm}
$
 \left.
    \begin{array}{ll}
\mathcal{P}_1 = \blacksquare \rightarrow \blacksquare \rightarrow \blacklozenge  \\
\mathcal{P}_2 = \blacksquare \rightarrow \blacktriangle\rightarrow \blacklozenge  \\
\mathcal{P}_3 = \blacksquare \rightarrow \blacktriangle\rightarrow \bigstar \rightarrow \blacklozenge
    \end{array}
\right \}\mathcal{E}_{\mathcal{P}}
$ \\ \\

$F((\blacksquare,\blacklozenge);\mathcal{E}_{\mathcal{P}}) = \beta_0 + \sum_{\mathcal{P} \in \mathcal{E}_{\mathcal{P}}} \beta_{\mathcal{P}} \, \text{PCRW}(\mathcal{P})$ \\
$\phantom{F((\blacksquare,\blacklozenge);\mathcal{E}_{\mathcal{P}})} \sim 
\text{PCRW}(\textcolor{red}{E_c})$
\vspace{4.5cm}
\end{tabular}
\label{IllDefc}
}
\vspace{-3.3cm}
\end{tabular}
\caption{\textbf{(a)} Example of HIN composed of multiple node types, represented by diverse shapes, an multiple link types. Nodes are already grouped by shapes. 
\textbf{(b)} Its associated network schema composed by five nodes and twenty links.
Each node corresponds to a set of nodes in the corresponding HIN. In the same way, each link is a set of links in the corresponding  HIN. See for instance the paths and meta-path of length two in blue $\blacksquare \rightarrow \blacktriangle\rightarrow \bigstar$; the blue paths are said to satisfy the blue meta-path. 
\textbf{(c)} Illustration of the problem statement. 
For each pair of nodes in ($\blacksquare$,$\blacklozenge$), there is possibly a link connecting them. The link weight is approximated by a linear combination of the path-constrained random walk results i.e., probability distributions of being at a particular node. 
Roughly speaking, the probabilities resulting from the random walk constrained by the target meta path $\textcolor{red}{E_c} = \blacksquare \rightarrow \blacklozenge$, denoted by  PCRW$(\textcolor{red}{E_c})$, are expressed as a linear combination $F$ of probabilities resulting from the random walks constrained by three different meta-paths $\mathcal{P}_1 = \blacksquare \rightarrow \blacksquare \rightarrow \blacklozenge$, $\mathcal{P}_2 = \blacksquare \rightarrow \blacktriangle\rightarrow \blacklozenge$, $\mathcal{P}_3 = \blacksquare \rightarrow \blacktriangle\rightarrow \bigstar \rightarrow \blacklozenge$, denoted by PCRW$(\mathcal{P}_1)$, PCRW$(\mathcal{P}_2)$ and PCRW$(\mathcal{P}_3)$ respectively, whose real-valued coefficients are $\beta_{\mathcal{P}_1}, \beta_{\mathcal{P}_2}$ and $\beta_{\mathcal{P}_3}$ respectively plus a possible independent term $\beta_0$, 
that is to say $F((\blacksquare,\blacklozenge);\mathcal{E}_{\mathcal{P}}) = \beta_0 + \sum_{\mathcal{P} \in \mathcal{E}_{\mathcal{P}}} \beta_{\mathcal{P}}\, \text{PCRW}(\mathcal{P})$. One can see that other meta-paths exist between nodes in ($\blacksquare$,$\blacklozenge$). The problem is to identify the ``best'' $\mathcal{E}_{\mathcal{P}}$ and a linear function $F$ with respect to PCRW$(\textcolor{red}{E_c})$.
}
\label{IllDef}
\end{figure*}

In this section, we present some basic concepts of weighted HIN useful for the following and define the ``weight recovery'' problem. Fig. \ref{IllDef} illustrates this section. 

\begin{definition}[Weighted directed multigraph]
A weighted directed mutligraph is a 5-tuple $G :=(V,E,w,\mu_s,\mu_t)$ with $V$ the node set, $E$ the link set, $w: E \rightarrow  \mathbb{R}^+$ the function that assigns to each link a real non negative weight, $\mu_s: E \rightarrow V$ the function that assigns to each link a source node, $\mu_t: E \rightarrow V$ the function that assigns to each link a target node.
\end{definition}

This concept allows us to introduce the definition of HIN which basically is a weighted directed multigraph with multiple node and link types.

\begin{definition}[Heterogeneous Information Network]
A HIN $H := (G,\mathcal{V},\mathcal{E},\phi,\psi)$ is a weighted directed multigraph $G$ along with $\mathcal{V}$ the node type set, $\mathcal{E}$ the link type set, $\phi: V \rightarrow \mathcal{V}$ the function that assigns a node type to each node and $\psi: E \rightarrow \mathcal{E}$ the function that assigns a link type to each link such that if two links belong to the same link type, the two links share the same starting and target node type i.e.,
$\forall\, e_1, e_2 \in E, \, \big( \psi(e_1) = \psi(e_2) \big) \Rightarrow \big( \phi(\mu_s(e_1)) = \phi(\mu_s(e_2)) \, \wedge \, \phi(\mu_t(e_1)) = \phi(\mu_t(e_2)) \big)$.
\end{definition}

Fig.\ref{IllDefa} illustrates such a network composed of five node types and twenty link types. However, disentangling the different entities present in the HIN is not necessarily a trivial task. Indeed, it sometimes takes a broader view of the system in question to describe it. For that purpose, the concept of network schema i.e. the meta level description of the HIN, is proposed. In simple terms, this corresponds to the graph defined over the node and link types of the associated HIN. It is represented in Fig.\ref{IllDefb}.  

\begin{definition}[HIN Schema]
Let H be a HIN. The schema $T_H$ for $H$ is a directed graph defined on the node types $\mathcal{V}$ and the link types $\mathcal{E}$ i.e., $T_H := (\mathcal{V},\mathcal{E},\nu_s,\nu_t)$ with $\nu_s : \mathcal{E} \rightarrow \mathcal{V} : E^* \mapsto \nu_s(E^*) := \phi \big(\mu_s (e) \big)$ the function that assigns to each link a source node and $\nu_t : \mathcal{E} \rightarrow \mathcal{V} : E^* \mapsto \nu_t(E^*) := \phi \big(\mu_t (e) \big)$ the function that assigns to each link a target node, where $e \in E$ such that $\psi(e) = E^*$
\end{definition}

Note that we can effectively take any such element $e$ since $\{e \in E \, \vert \, \psi(e) = E^* \}$ is the equivalence class of any of its elements, with the equivalence relation ``has the same type of''. By definition of HIN, it is sufficient to take one member of the equivalence class to know the node types that the link type $E^*$ connects.

Two entities in a HIN can be linked via different paths and these paths have different semantics. These paths can be defined as meta-paths as follows \cite{shi2016survey}.

\begin{definition}[Meta-path]
A meta-path $\mathcal{P}$ of length $n \in \mathbb{N}$ is a sequence of node types $V_0, \cdots, V_n \in \mathcal{V}$ linked by link types $E_1, \cdots, E_n \in \mathcal{E}$ as follows: $\mathcal{P} = V_0 \, \xrightarrow{E_1} \, V_1 \, \cdots \,  V_{n-1} \, \xrightarrow{E_{n}} \, V_{n}$ which can also be denoted as $\mathcal{P} = E_1 E_2 \cdots E_n$.
\end{definition}

Given a meta-path $\mathcal{P} = V_0 \,\, \xrightarrow{E_1} \,\, V_1 \,\, \cdots \,\,  V_{n-1} \,\,  \xrightarrow{E_n} \,\, V_n$ and a path $P = v_0 \,\, \xrightarrow{e_1} \,\, v_1 \,\, \cdots \,\,  v_{n-1} \,\,  \xrightarrow{e_n} \,\, v_n$, if $\forall \, i  \in \{0, ..., n \}, \ \phi(v_i)=V_i, \ \forall \, i  \in \{1, ..., n \}, \ \mu_s(e_i) = v_{i-1}, \, \mu_t(e_i)=v_i$ and $\psi(e_i) = E_i$, then path $P$ satisfies meta-path $\mathcal{P}$ and we note $P \in \mathcal{P}$. Hence, a meta-path is a set of paths.

In Fig.\ref{IllDefb}, an example of meta-path is $\blacksquare \rightarrow \blacktriangle\rightarrow \bigstar$, in blue, in the network schema. Blue paths in the HIN in Fig.\ref{IllDefa} are said to satisfy this meta-path since each one of their segments respects the aforementioned conditions.

\begin{problem}[Weight recovery]
Let be a HIN $H=(G,\mathcal{V},\mathcal{E},\phi,\psi)$, with $G=(V,E,w,\mu_s,\mu_t)$ a directed weighted multigraph, and a target link type $E_c$ between two node types. The ``weight recovery problem'' is to find a set of relevant meta-paths $\mathcal{E}_{\mathcal{P}}$ and a linear function $F$ of probabilities resulting from random walks constrained by these meta-paths that best quantifies, for each pair of nodes in $H$, the strength of their connection via $E_c$. 
\label{Problem1}
\end{problem}

\section{METHOD}
\label{Method}

We present our method for solving Problem \ref{Problem1} in three steps. 
Consider a HIN and let us denote by $E_c$ the target link type defined between $V_0$ and $V_n$.
We consider a meta path $\mathcal{P} = V_0 \, \xrightarrow{E_1} \, V_1 \, \cdots \,  V_{n-1} \,  \xrightarrow{E_n} \, V_n$ different from $E_c$. There may be repetitions in this sequence of nodes and links. Let us introduce the notation $\mathcal{P} \equiv \mathcal{P}^{0,n}$ and let us denote by $\mathcal{P}^{a,b}$ the truncated meta path of $\mathcal{P}$ from node type $V_a$ to $V_b$.

\subsection{Path-Constrained Random Walk.}
\label{Subsection_PCRW}

Let $X_i \in V_i$ be a random variable representing the position of a random walker  in the set $V_i$. A random walk starting from $X_0$ constrained by the meta-path $\mathcal{P}$ corresponds to a discrete-time Markov chain i.e., a sequence of random variables $X_0,\, X_1, ...,\, X_n$ with the Markov property: $\forall \, i \in \{0,...,n \}, \, \forall \, (v_0,...,v_n) \in V_0 \times ... \times V_n, $
\begin{equation*}
\mathbb{P}(X_i=v_i \, \vert \, X_{i-1}=v_{i-1}, ..., X_0 = v_0) = \mathbb{P}(X_i=v_i \, \vert \, X_{i-1}=v_{i-1}).
\end{equation*}

Here, since there may be more than one link type between two node types, we introduce de notation $\mathbb{P}((X_i=v_i \, \vert \, X_{i-1}=v_{i-1}) \, \vert \, \mathcal{P}^{i,i+1}) =: \mathbb{P}((v_i \, \vert \, v_{i-1}) \, \vert \, \mathcal{P}^{i,i+1}) = \mathbb{P}((v_i \, \vert \, v_{i-1}) \, \vert \, E_i)$ to emphasize the fact that the random walk is constrained by the meta-path $\mathcal{P}$. 
This means that for a walker to reach $v_i$ from $v_{i-1}$, he has to follow only links of type $E_i \equiv \mathcal{P}^{i,i+1}$. 
The probability $\mathbb{P}((v_i \, \vert \, v_{i-1}) \, \vert \, E_i)$ thus defined is computed as 
\begin{equation*}
\mathbb{P}((v_i \, \vert \, v_{i-1}) \, \vert \, E_i) = 
\frac{w_{E_i}(v_{i-1},v_i)}{\sum_k w_{E_i}(v_{i-1},v_{k})}
\end{equation*}
\noindent where $w_{E_i}(v_j,v_k)$ is the link's weight of type $E_i$ between nodes $v_j$ and $v_k$.

Thenceforth, given $v_n \in V_n$ and $v_0 \in V_0$, the probability of reaching $v_n$ from $v_0$ following the meta path $\mathcal{P}$, denoted by $\mathbb{P}((v_n \vert v_0) \, \vert \, \mathcal{P})$, is simply defined by the random walk starting at $v_0$ and ending at $v_n$ following only paths satisfying $\mathcal{P}$. 
This conditional probability may be expressed recursively by means of the law of total probability
\begin{align}
\mathbb{P}((v_n \vert v_0) \, \vert \, \mathcal{P})
&= \sum \limits_{v_{n-1} \in V_{n-1}}
\bigg[
\mathbb{P}\bigg((v_n \vert v_{n-1}) \, \vert \, E_n \bigg) \, \nonumber \\
& \phantom{\hspace{1cm}} \times 
\mathbb{P}\bigg((v_{n-1} \vert v_0) \, \vert \, \mathcal{P}^{0,n-1}  \bigg) 
\bigg] 
\nonumber \\ 
&= \sum \limits_{v_{n-1} \in V_{n-1}} 
\bigg[
\frac{w_{E_n}(v_{n-1},v_{n})}{\sum_k w_{E_n}(v_{n-1},v_{k})} \, \nonumber \\
& \phantom{\hspace{1cm} } \times
\mathbb{P}\bigg((v_{n-1} \vert v_0) \, \vert \, \mathcal{P}^{0,n-1}  \bigg)
\bigg]
\label{PCRW}
\end{align} 
\noindent with $\mathbb{P}((v_1 \vert v_0) \vert \mathcal{P}^{0,1})=w_{E_1}(v_0,v_1) / \sum_k w_{E_1}(v_0,v_k)$ the basis of recurrence. 
In the following, we use the notation PCRW($\mathcal{P}$) to denote the column vector of such conditional probabilities $\mathbb{P}((v_n \vert v_0) \, \vert \, \mathcal{P})$, $\forall \, v_0, \, v_n$ i.e., PCRW($\mathcal{P}$) = $[\mathbb{P}((v_0 \vert v_0) \, \vert \, \mathcal{P}), \mathbb{P}((v_1 \vert v_0) \, \vert \, \mathcal{P}), ...., \mathbb{P}((v_n \vert v_n) \, \vert \, \mathcal{P})]^{\rm {T}}$.

For instance, in the HIN in Fig.\ref{IllDefa}, the probability for a walker to reach the green star \textcolor{vert}{$\bigstar$} from the green square \textcolor{vert}{$\blacksquare$} following the meta-path $\blacksquare \rightarrow \blacktriangle\rightarrow \bigstar$ equals 5/12.

Note that we forbid the walker to return to the initial node on the penultimate step of the walk i.e., if $V_{n-1}=V_0$, the sum in eq. \eqref{PCRW} only holds for all $v_{n-1} \neq v_0$. 
It prevents us from using what we are looking for to find what we are looking for.

\begin{remark}[Hole nodes]
It is possible that a node $v_i \in V_i$ is not connected to any node $v_j \in V_j$ by the link type $E_{ij}$ and thus, the transition probability is not defined. To overcome this problem, we provide each set $V_k$ with a hole node $h_k$ on which point all the disconnected nodes. Plus, all the holes are connected with each other and holes cannot point to another node (i.e., no hole node). 
Formally, $\forall \, V_k \in \mathcal{V}, \, V^h_k := V_k \cup \{ h_k \}$. $\forall \, E_{ij} \in \mathcal{E}$, if $w_{E_{ij}}(v_i,v_j)=0, \, \forall \, v_j \in V_j$ then $w_{E_{ij}}(v_i,h_j)=1$, otherwise $w_{E_{ij}}(v_i,h_j)=0$. Furthermore, $\forall \, E_{ij} \in \mathcal{E}$, $w_{E_{ij}}(h_i,h_j)=1$ and $\sum_{v_j \in V_j} w_{E_{ij}}(h_i,v_j)=0 $.
In this fashion, transition probabilities are always well defined. 
\end{remark}

\subsection{Linear Regression Model.}
Since $H$ is a HIN, multiple types of links can connect the nodes. Hence, there is no reason to restrict ourselves to a single meta path to compute the reachability of one node from another. 
As a result, the similarity between $v_n$ and $v_0$ is defined by several path-constrained random walk results combined through a linear regression model of the form
\begin{equation*}
F((v_n \vert v_0) \, \vert \, \mathcal{E}_\mathcal{P})
:=
\beta_0 + 
\sum \limits_{\mathcal{P} \in \mathcal{E}_\mathcal{P}}
\beta_{\mathcal{P}} \ \mathbb{P}((v_n \vert v_0) \, \vert \, \mathcal{P})
\end{equation*}
\noindent where $\mathcal{E}_\mathcal{P}$ is the set of selected meta-paths and the vector {\boldm $\mathbf{\upbeta}$} $:= [\beta_0, \beta_1, \cdots, \beta_{\vert \mathcal{E}_\mathcal{P} \vert}]^{\rm {T}}$ is real-valued coefficients. 
The coefficients stress the contribution of each meta-path in the final similarity score $F((v_n \vert v_0) \, \vert \, \mathcal{E}_\mathcal{P})$.
Since the components of {\boldm $\mathbf{\upbeta}$} are not confined in [0,1] and do not sum to 1, $F$ is a real-valued function whose image is neither confined in [0,1].

Now, we have a linear regression problem since we want to recover the exact link weights with respect to $E_c$. The dependant variable is thus PCRW$(E_c)$ whilst the predictors are PCRW$(\mathcal{P})$, $\mathcal{P} \in \mathcal{E}_\mathcal{P}$. 
The choice of linear model is simply motivated by its interpretability in our particular case. 
Given example node pairs and their link weights, {\boldm $\mathbf{\upbeta}$} is estimated by the least squares method which is appreciated for its applicability and simplicity. In formulae with $\mathds{1}$ the column vector whose entries are 1: \\

\begin{footnotesize}
$
\begin{blockarray}{c}
    \text{{\tiny PCRW}}(E_c) \\
    \downarrow \\ 
    \begin{block}{[c]}
	\\ \\ \text{{\tiny PCRW}}(E_c) \\ \\ \\
    \end{block} \\
\end{blockarray}
$
=
$
\begin{blockarray}{cccc}
    \mathds{1} & {\tiny\text{PCRW}(\mathcal{P}_0)} & \dots & {\tiny\text{ PCRW}(\mathcal{P}_{\vert \mathcal{E}_{\mathcal{P}} \vert})}\\
    \downarrow & \downarrow &  & \downarrow \\ 
    \begin{block}{[cccc]}
         &    &      & \\
         &    &      & \\
         \BAmulticolumn{4}{c}{\text{PCRW}(\mathcal{E}_{\mathcal{P}})}  \\
         &    &      & \\
         &    &      & \\
    \end{block} \\
\end{blockarray}
\begin{blockarray}{c}
     \\
     \\ 
    \begin{block}{[c]}
     \\ \\  {\Large \boldsymbol{\beta}}   \\ \\ \\
    \end{block} \\
\end{blockarray}
$
+
$
\begin{blockarray}{c}
     \\
     \\ 
    \begin{block}{[c]}
     \\ \\  {\LARGE \text{$\epsilon$}}  \\ \\ \\
    \end{block} \\
\end{blockarray}$ \\
$ \hspace*{1.8cm}
= 
\begin{bmatrix}
{\large \text{ $ F(\mathcal{E}_\mathcal{P}) $ }}
\end{bmatrix}
+
\begin{bmatrix}
{\LARGE \text{$\epsilon$}}
\end{bmatrix}
$ \\
\end{footnotesize}

\noindent and we choose {\boldm $\mathbf{\hat{\upbeta}}$} such that the residual sum of squares RSS $=\epsilon^T \epsilon = \Vert \epsilon \Vert^2$ is minimized i.e., {\boldm $\mathbf{\hat{\upbeta}}$} = $\big( \text{PCRW}(\mathcal{E}_{\mathcal{P}}) ^{\rm {T}} \, \text{PCRW}(\mathcal{E}_{\mathcal{P}})  \big)^{-1} \, \text{PCRW}(\mathcal{E}_{\mathcal{P}}) ^{\rm {T}} \, \text{PCRW}(E_c) $.

\subsection{Forward Selection Procedure.}
In order to determine the set $\mathcal{E}_{\mathcal{P}}$, we use the forward selection with $p$-value and $r^2$ criteria. This is a greedy approach but very simple and intuitive. The $p$-values are used to test the significance of each predictor. Given the hypothesis $H_0: \beta = 0$ against the hypothesis $H_1: \beta \neq 0$, the $p$-value $p$ is the probability, under $H_0$, of getting a statistics as extreme as the observed value on the sample. We reject the hypothesis $H_0$, at the level $\alpha$, if $p \leq \alpha$ in favor of $H_1$. Otherwise, we reject $H_1$ in favor of $H_0$.
Conversely, the $r^2$ score is used to test the quality of the entire model. It is the proportion of the variance in the dependent variable that is predictable from the predictors. 
Note that the $r^2$ = 1-RSS/TSS where TSS is the total sum of squares i.e., is the sum of the squares of the difference of the dependent variable and its mean. Hence, maximizing the $r^2$ is equivalent to minimizing the RSS.

So, given $k$ predictors or explanatory variables which are the probability distributions PCRW($\mathcal{P}_k$), the forward selection procedure works as follows
\vspace{-0.6em}
\begin{itemize}
\itemsep0em
\item Start with a null model i.e. no predictor but only an intercept. Typically, this is the average of the dependent variable;
\item Try $k$ linear regression models (i.e., models with only one predictor) and chose the one which gives the best model with respect to the criterion. In our case, the one that minimizes RSS or alternatively, the one that maximizes the coefficient of determination $r^2$;
\item Search among the remaining variables the one that, added to the model, gives the best result i.e., the higher $r^2$ such that all the variables in the model are significant i.e., their $p$-value is below the chosen threshold. Iterate this step until no further improvement.
\end{itemize}

\subsection{Validation}
Since we would like to use the regression model as a prediction model (i.e., not only a descriptive one), we use Monte Carlo cross-validation a.k.a. repeated random sub-sampling validation \cite{xu2001monte}.
Given a data set of $N$ points, the method simply splits them into a training subset $s_t$ and a test subset $s_v$. The model is then trained on $s_t$ and tested on $s_v$. This procedure is repeated multiple times and the results are then averaged over the splits.
Note that the results of Monte Carlo cross-validation tends towards those of leave-$p$-out cross-validation \cite{arlot2010survey} as the number of random splits tends to infinity.
The drawbacks of this method are the possibility that some observations may never be selected for training or on the contrary, may be used at each split. Plus, the results depend on the different random splits i.e., it displays Monte Carlo variation.
However, it has advantage (over $k$-fold cross validation \cite{arlot2010survey}) as the proportion of the split is independent of the folds (iterations). It means Monte Carlo allows to explore somewhat more possible partitions, though one is unlikely to get all of them since there exist ${\scriptsize \begin{pmatrix} N \\ s_t \end{pmatrix}}$ unique training subsets.

\begin{remark}[Division of a node type]
\label{Division_NT}
Given a HIN $H$ with $\mathcal{V} = \{ V_1, ..., V_k, ..., V_m \}$ the set of node types with $V_k = \{V_{k,1}, ..., V_{k,q} \}$, one can want to understand the ``role'' of each $V_{k,r}$. 
Let two node types $V_i$ and $V_j$ (not necessarily distinct) be the target node types and $\mathcal{E}_{\mathcal{P}}$ the set of meta-paths.
Plus, let $V_i$ and $V_j$ be linked by a specific meta-path including the node type $V_k$, namely, $\mathcal{P} =  V_i \,\,  \cdots \,\, \xrightarrow{e_k} \,\, V_k \cdots \,\, \xrightarrow{e_j} \,\, V_j$ with $\mathcal{P} \in \mathcal{E}_{\mathcal{P}}$. 
We can thus construct $q$ subsets $S_{i,r} = \{ v_i \in V \, \vert \, \phi(v_i) \in V_i \wedge \exists \, P = v_i \,\,  \cdots \,\, \xrightarrow{e_k} \,\, v_{k,r} \cdots \,\, \xrightarrow{e_j} \,\, v_j \}$ and $q$ subsets $S_{j,r} = \{ v_j \in V \, \vert \, \phi(v_j) \in V_j \wedge \exists \, P = v_i \,\,  \cdots \,\, \xrightarrow{e_k} \,\, v_{k,r} \cdots \,\, \xrightarrow{e_j} \,\, v_j \}$ ($r=1,..,q$) such that with $v_{k,r} \in V_{k,r} \subseteq V_k$ ($v_j \in V_j$ and $v_i \in V_i$ resp.) and $P \in \mathcal{P}$.
We can thus build $q$ linear regression models: one for each HIN $H_r$ formed from the node set $\big\{v \in V \, \vert \, \phi(v) \in \mathcal{V} \setminus \{ V_k, V_i, V_j \} \big\} \cup\{ S_{i,r}, S_{j,r} \} $ with meta-paths $\mathcal{E}_{\mathcal{P}} \setminus \mathcal{P}$.
Analysing the vector {\boldm $\mathbf{\hat{\upbeta}}$} of each final model can bring some insight about the ``role'' of each $V_r$.
\end{remark}

\section{EXPERIMENTS}
\label{Experiments}

We test the proposed methods on two real-worls data sets. The first one, related to FIFA WorldCup 20104 Twitter data, allows us to perform tests between target nodes with different types. The task consists in recovering the user-hashtag frequency. The second data set, related to bibliographic data, focuses on target nodes of the same types and tackles the problem of co-authorship.

\begin{figure*}[h]
\centering 
\begin{minipage}[c]{0.42\linewidth}
\includegraphics[scale=1.3]{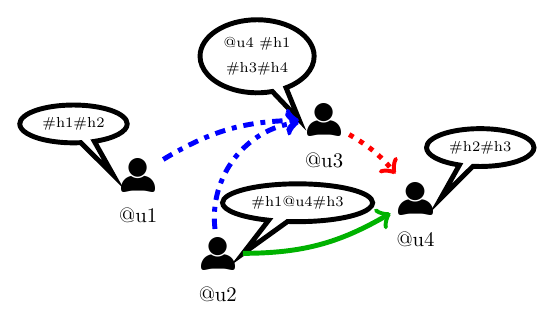}
\end{minipage} 
\begin{minipage}[c]{.1\linewidth}
\centering
\includegraphics[scale=1]{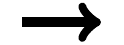}
\end{minipage} 
\begin{minipage}[c]{.46\linewidth}
\centering
\begin{tabular}{cc}
\includegraphics[scale=1.1]{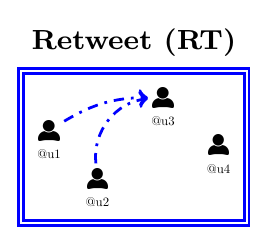}
&
\includegraphics[scale=1.1]{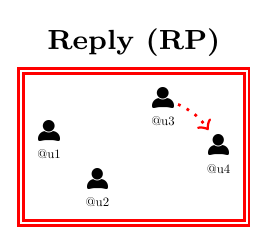} \\
\includegraphics[scale=1.1]{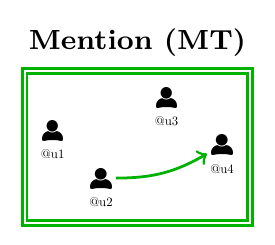}
&
\includegraphics[scale=1.1]{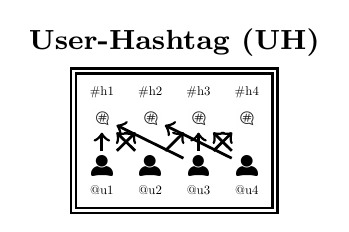}
\end{tabular}
\end{minipage}

\caption{Illustration of the construction of graphs based on Twitter interactions where four users interact with each other through three types of interactions: retweet, reply and mention; and write some hashtags in their post. The underlying HIN is such that $\mathcal{V}=$\{users, hashtags\} and $\mathcal{E}=$\{RT, RP, MT, UH\}. The four graphs associated to the types of actions are displayed separately for convenience.}
\label{TwitterInteraction}
\end{figure*}

\subsection{FIFA WorldCup 2014 Twitter data.}
\label{Experiments_Different}

We present the data set on which we test the proposed method as well as the construction of the resulting graphs. Then, we report our results concerning different tests namely, the importance of meta path length, a description task and finally a recovery task. 

\subsubsection{Data Set Description and Setup.}

The data we use is a set of tweets collected from Twitter during the Football World Cup 2014. This period extents from June 12 to July 13, 2014. 
Twitter allows multiple kinds of interactions between its users. Here, we consider retweet (RT), reply (RP) and mention (MT) actions plus the fact of posting hashtags (UH). 
The RT relationship means that a user broadcasts a tweet previously posted by another user. The RP action is simply a response tweet to another user in connection with her previous tweet. The last action considered here is the MT action. This happens when a user mentions explicitly another user in her post.

Based on these actions, we construct a HIN with two node types $\mathcal{V}=$\{users, hashtags\} and four edge types $\mathcal{E}=$\{RT, RP, MT, UH\} as illustrated in Fig. \ref{TwitterInteraction}. 
Each node represents a user or a hashtag. We create a link from $u_1$ to $u_2$ if $u_1$ retweets, replies (to) or mentions $u_2$ and the weight of the link correspond to the number of times $u_1$ performs the specific action towards $u_2$ during the whole world cup. For the user-hashtag graph, a link exists between $u$ and $h$ if $h$ appears in $u$'s post and the weight of the link corresponds to the number of times $u$ post $h$ during the whole world cup. Note that we exclude hashtags present in the retweeted posts since in these cases, users do not write them themselves. Furthermore, considering them would provoke a trivial correlation between UH and RT-UH. All graphs are directed and weighted. 

The data set contains 13,826 users and 14,392 hashtags.
The RT graph is composed of 6,069 nodes and 19,495 links, the RP graph is composed of 8,560 nodes and 11,782 links and the MT graph is composed of 11,782 nodes and 60,506 links.
Note that Pearson coefficient between the stochastic matrices rises to 0.1776, 0.6783 and 0.4286 for RT/RP, RT/MT et RP/MT respectively. Thus, the retweet and mention relationships are clearly correlated which may cause some problems for the proposed method, as we shall see, since it is well known that least squares method is sensitive to that.
Since the data is related to the world cup, the most used hashtags of bipartite users-hashtags graph UH are those referring to the 32 countries involved in the final phase as well as those referring directly to the event (\#WorldCup2014, \#Brazil, \#Brasil2014, \#CM2014, ...). The semi finalists have the greatest in-strength (in-strength of the node $j$ is $s^{in}_j = \sum_{i} w_{ij}$).

\subsubsection{Results.}
We apply the proposed method to find if the hashtags posted by users (UH) can be explained by other relations (RT, RP, MT and their combinations). For instance, given a user $u$, explaining UH by RT-UH and MT-RP-UH means that the hashtags posted by $u$ are, to some extent, a combination of those posted by the users retweeted by $u$ and those posted by the users who received a response from users mentioned by $u$. In other words, we try to understand if, in the case of the football World Cup 2014, the probability that users post hashtags can be explained by the relations these users have with other users and the probability that these latter have to post these hashtags. 

\textbf{Meta-Paths of Length 2.}
We test linear regression models with all the possible combinations of meta-paths of length 2 (see Table \ref{RegBase}). This test allows a first glimpse of the contribution of the simplest predictors. 
\begin{table}[!b]
\centering 
\begin{tabular}{ccccc}
\hline
\rowcolor{gray!50}
Mod. & Meta-Path & Coef. & $p$-values & $r^2$ \\
\hline
A0 & \multicolumn{3}{c}{Average : 1.8704e-05} & 0.2992 \\
\hline
\rowcolor{gray!25}
A1 & RT-UH & 0.6273 & - & 0.3594 \\
\hline
B1 & RP-UH & 0.4291 & - & 0.2289 \\
\hline
\rowcolor{gray!25}
C1 & MT-UH & 1.0289 & - & 0.4606 \\
\hline
\multirow{2}{*}{A2} & RT-UH & 0.5795 & 0.0062 & \multirow{2}{*}{0.6116} \\
& RP-UH & 0.3957 & 0.0105 & \\
\hline
\rowcolor{gray!25}
& RT-UH & -0.3578 & 0.0612 & \\
\rowcolor{gray!25}
\multirow{-2}{*}{B2} & MT-UH & 1.4534 & 0.0087 & \multirow{-2}{*}{0.5943} \\
\hline
\multirow{2}{*}{C2} & RT-UH & 0.0051 & 0.0138 & \multirow{2}{*}{0.6111} \\
& MT-UH & 0.9391 & 0.0057 & \\
\hline
\rowcolor{gray!25}
& RP-UH & -0.1283 & 0.0791 & \\
\rowcolor{gray!25}
& RP-UH & 0.0791 & 0.0113 & \\
\rowcolor{gray!25}
\multirow{-3}{*}{A3} & MT-UH & 1.1466 & 0.0111 & \multirow{-3}{*}{0.6818} \\
\hline
\end{tabular}
\caption{Coefficients and $p$-values for linear regressions whose regressors correspond to meta-paths of length 2 in order to explain the user-hasthag distribution (UH). Model A0 corresponds to the null model: no predictor but one intercept that is the average of the explained variable.}
\label{RegBase}
\end{table}
First, the more the predictors, the better the value of $r^2$. It thus could be tempting to consider them all. Nevertheless, it does not mean that all predictors are significant.
Indeed, the analysis of the coefficients and $p$-values makes it possible to realize the correlation of some variables. In models B2 and A3, the RT-UH and MT-UH meta-paths are both present. However, the $p$-value associated to RT-UH is greater than 0.05 which states that we accept the null hypothesis for this predictor. This could be a consequence of the correlation between RT-UH and MT-UH.

In summary and as it can be seen in Table \ref{RegBase}, the best model according to the $r^2$ and the $p$-values with threshold $\alpha=0.05$ would be the model A2 whose predictors are RT-UH and RP-UH. The gain in the $r^2$ with respect to any other model with 1 regressor (and so simpler model) is worth it i.e., important $r^2$ improvement and not really more complexity added.
This means that, for a given user, the hashtags she posts can be explained by the hashtags posted by the users she retweets with a contribution of 0.5795 and the users she replies to with a contribution of 0.3957. This model accounts for 61.16\% of the variance.

\textbf{Importance of Meta Path Length.}
This subsection looks at the length of the meta-paths for a given link type. 
More specifically, we compute, for each link type, the $r^2$ score when the only predictor is associated to a random walk of length $l=1,...,10$ repeating the same link type. For instance, for $l=2$ and the retweet action, the predictor will be RT-RT-UH representing the hashtags posted by people who are retweeted by people who are themselves retweeted. 
Intuitively, the importance of a meta-path decreases with its length ($=l+1$) since considering longer meta-paths means considering more extended neighborhoods, hence the information is more diffused. By way of illustration, the walker can attain a lot of nodes with some of them really far from the starting node.

\begin{figure}[h]
\captionsetup[subfigure]{position=top, labelfont=bf,textfont=normalfont,singlelinecheck=off,justification=raggedright}
\centering
\subfloat[]{\includegraphics[scale=0.48]{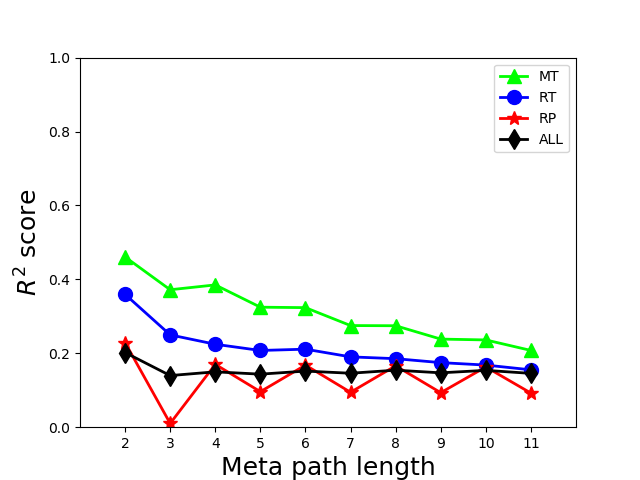}\label{IterNumVara}}

\subfloat[]{\includegraphics[scale=0.48]{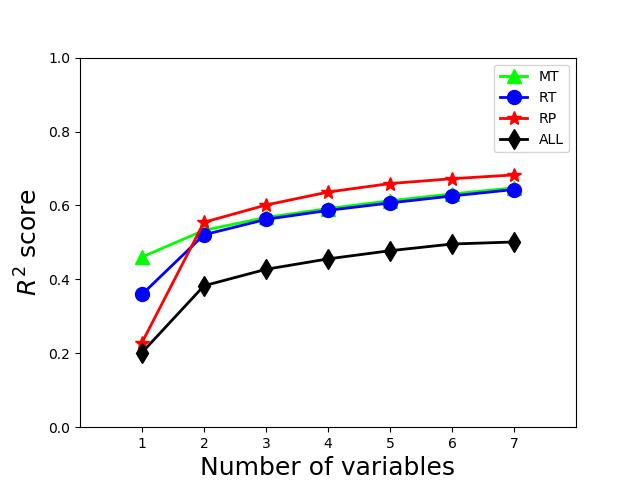} \label{IterNumVarb}}
\caption{(\textbf{a)} Linear regression $r^2$ scores with one predictor associated to a meta-path whose length varies between 2 and 11. 
\textbf{(b)} Linear regression $r^2$ scores according to the number of meta-paths of the same link type (see main text for explanation).}
\label{IterNumVar}
\end{figure}

This is corroborated in Fig.\ref{IterNumVara} where we can see a tendency to decrease with respect to the meta-path length.
Each link type brings a different quantity of information and the MT type is the more informative for our purpose.

\begin{figure}[h]
\centering
\includegraphics[scale=1]{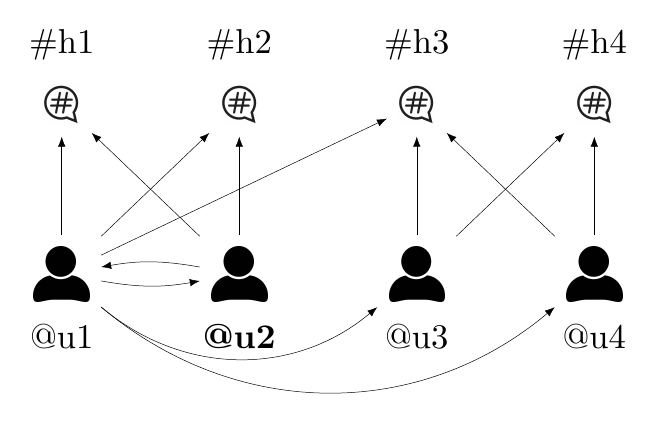}
\caption{Typical example of reply case focused on user $u_2$. The hashtags posted by $u_2$ are $h_1$ and $h_2$. The probabilities resulting from the random walk UH starting from $u_2$ are then $[1/2,1/2,0,0]^T$. For meta-path of length 2, a walker starting from $u_2$ following meta-path RP-UH has to go, with probability 1, to $u_1$ and then to $h_1$, $h_2$ and $h_3$. The resulting probabilities are $[1/3,1/3,1/3,0]^T$.
Now, for meta-path of length 3, the walker can not return to $u_2$ after being on $u_1$: he has to go to $u_3$ or $u_4$. But since these latter are not in connection with $u_2$ via the reply action, their hashtags are more different. This time, the probabilities are $[0,0,1/2,1/2]^T$, which is far from those obtained with UH: $[1/2,1/2,0,0]^T$. Consequently, the $r^2$ is really low (in this case, it is null). However, for meta-path of length 4, the walker can return to $u_2$ after being on $u_1$ so in the next step (the third step), the walker can only jump to $u_1$ who is a direct neighbor of $u_2$. The rationale  is the same for longer meta-paths: for even lengths, the walker is not affected by the restriction on the penultimate step of the walk while for odd lengths, it has huge importance.}
\label{Fig_ExplainUH}
\end{figure}

Plus, this analysis exposes a characteristic of the reply dynamics: most of the time, the replies involved only two people \cite{macskassy2012study}. This is reflected through the oscillations of the reply scores in Fig. \ref{IterNumVara}. The scores associated to odd length random walks are low since the walker is forbidden to return to the initial node on the penultimate step of the walk (see Fig. \ref{Fig_ExplainUH}).

We also draw in black the $r^2$ scores when we do not differentiate the link types (ALL) i.e., all the link weights between to nodes are aggregated. This score is below the average score of the three specific link types. One can see that just taking the mention or retweet type is more informative than the aggregation which reinforces the relevance of differentiating the link types.

Fig. \ref{IterNumVarb} shows $r^2$ scores when we combine variables of different lengths related to the same link type in the model. 
Actually, the $r^2$ associated to $n$ number of variables is related to the model whose predictors are all meta-paths of length smaller or equals to $n+1$ and whose the steps except the last are in the same type of links. 
For instance, for 3 variables, the predictors are RT-UH, RT-RT-UH and RT-RT-RT-UH (for the RT case). 
Again, the more the variables, the better the score. Also, the increase is not linear; the best improvement happens when we combine length-1 and length-2 variables which indicates the need to consider them together.
We can also observe that scores given by the RT and MT types are really similar when considering more than two variables while there is a clear difference in the $r^2$ score for single variable. It means that their respective combinations have the same result in term of $r^2$ although the underlying semantics are different.
Once again, the $r^2$ score for the aggregation is shown and is far below the other scores. This indicates that it is important to distinguish the types of links.

Since it is often desirable to keep a model simple both in term of interpretability and computation time, there is a trade-off between the highest possible $r^2$ and the cost to attain it. 
The tests here performed tend to show that considering too long as well as too many meta-paths is not necessarily useful in our case. Indeed, the gain in the $r^2$ is not worth it considering the complexity it brings.

\textbf{Forward Linear Regression for Data Description.}
We apply the proposed algorithm on the entire data set with a threshold $\alpha = 0.05$ for $p$-values. As a reminder, the procedure stops when there is no longer possible to improve the $r^2$ by adding significant regressors. Since the length of meta-paths is unbounded, the set of possible meta-paths is infinite. Here, the $k$ potential predictors are those of length less than or equal to 4. This is motivated by the test performed in the previous subsection. In addition, the semantics of longer paths are less clear than shorter paths.

\begin{figure}[t]
\centering 
\includegraphics[scale=0.54]{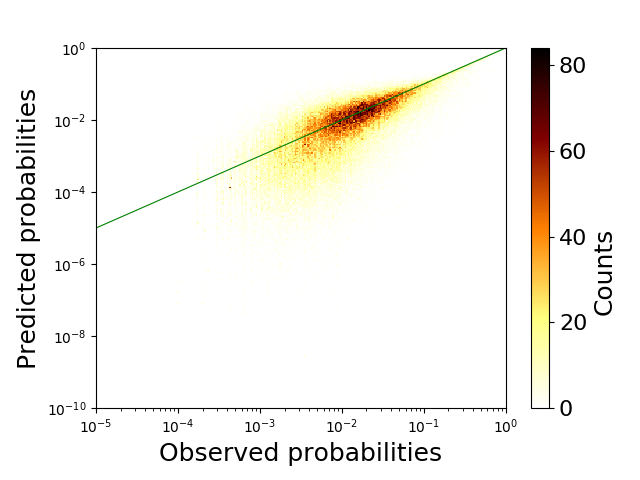}
\caption{Density plot of observed versus estimated values for the model 5. Green line represents the perfect matching between observed and estimated data.}
\label{ScatterPLot}
\end{figure}

Results are reported in Table \ref{StepwiseRes}. The final model thus obtained contains five predictors related to meta-paths whose length are no longer than 3 and no intercept. This regression model accounts for 71.29\% of the variance. To comfort the goodness of fit of the model, we plot in Fig. \ref{ScatterPLot} the density plot in log-log scale of the predicted probabilities versus the observed ones in the data. The green line represents the ideal case where predicted probabilities match observed ones. Most of the data points fall to this line which reinforces the use of a linear model.

\begin{table}[t]
\centering
\begin{tabular}{ccccc}
\hline
\rowcolor{gray!50}
Mod. & Meta-Path & Coef. & $p$-values & $r^2$ \\
\hline
0 & \multicolumn{3}{c}{Average: 1.8704e-05} & 0.2992 \\
\hline
\rowcolor{gray!25}
1 & MT-UH & 1.0289 & - & 0.4606 \\
\hline
\multirow{2}{*}{2} & MT-UH & 0.9391 & 0.0057 & \multirow{2}{*}{0.6112} \\
& RP-UH & 0.0052 & 0.0137 & \\
\hline
\rowcolor{gray!25}
& MT-UH & 0.8464 & 0.0062 & \\
\rowcolor{gray!25}
& RP-UH & 0.0335 & 0.0124 & \\
\rowcolor{gray!25}
\multirow{-3}{*}{3} & RT-RP-UH & 0.1077 & 0.0138 & \multirow{-3}{*}{0.6682} \\
\hline
\multirow{4}{*}{4} & MT-UH & 0.8114 & 0.0063 & \multirow{4}{*}{0.6947} \\
& RP-UH & 0.0362 & 0.0109 & \\
& RT-RP-UH & 0.0766 & 0.0142 & \\
& RP-MT-UH & 0.0676 & 0.0143 & \\
\hline
\rowcolor{gray!25}
& MT-UH & 0.1974 & 0.0094 & \\
\rowcolor{gray!25} 
& RP-UH & 0.5556 & 0.0146 & \\
\rowcolor{gray!25} 
& RT-RP-UH & 0.0650 & 0.0125 & \\
\rowcolor{gray!25} 
& RP-MT-UH & 0.1591 & 0.0160 & \\
\rowcolor{gray!25} 
\multirow{-5}{*}{5} & MT-RT-UH & 0.0074 & 0.0124 & \multirow{-5}{*}{0.7129} \\
\hline
\end{tabular}
\caption{Results of the forward stepwise linear regression.}
\label{StepwiseRes}
\end{table}

The best improvement with respect to $r^2$ comes with the addition of the second variable (see Mod. 2 of Table \ref{StepwiseRes}). 
The model with two predictors is actually a local extremum since the model with the best $r^2$ is the one with RT-UH and RP-UH predictors (see Table \ref{RegBase}). Although the difference is tenuous, this allows to point two weaknesses of the method: there is no guarantee of finding the best model and the order of the variable selection is important.
Note that the first two variables are part of the most direct relationships (meta-paths of length 2) which is intuitive: the direct neighborhood of a user shares common topics with her.
The last meta path included in the model (Mod. 5) provokes an important change in the other coefficients. This suggests this meta-path is either correlated to other meta-paths already present in the model or the presence of outliers i.e., 
observation which is ``distant'' from other observations. It is well known that ordinary least squares method is sensitive to that.

\textbf{Forward Linear Regression for Data Recovery.}
We validate the method by performing a task aiming to recover the weights of missing links. 
In other words, this part tries to answer to the question: is it possible to know, in a quantitative way, the way some people post some hashtags, knowing the way other people do ? 

\begin{figure}[h]
\centering 
\includegraphics[scale=0.5]{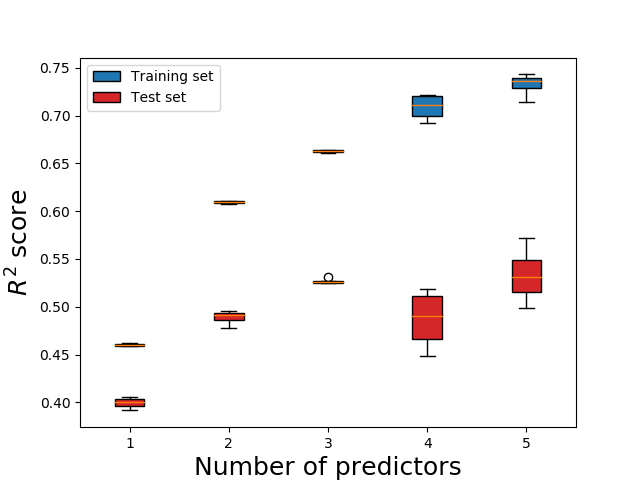}
\caption{Boxplot of the $r^2$ scores of training sets and test sets. The training set scores increase with the number of predictors in the model while for the testing set, the scores seem to reach a threshold.}
\label{PredTask}
\end{figure}

We perform Monte Carlo cross-validation with 80\% of the users as the training set and obtain the vector {\boldm $\mathbf{\upbeta}$} for them. Then, we use it on the testing set i.e. the remaining 20\% and compute the $r^2$ associated to each model. We proceed to ten splits i.e., we create ten training sets. 
The final models do not include the same variables as before. Not surprisingly, it depends on the 80\% selected. The number of predictors is five or six. 
Nevertheless, whatever the training set, the meta-path MT-UH is always the first predictor to be selected. After, there is no more consensus on the second regressor but the RP-UH and RT-RP-UH always compete for the second place. Again, it is not surprising to obtain the RP-UH meta-path since, for a user, it is related to one of the closest neighbors with respect to our graph construction and very weakly correlated to the MT-UH meta-path already present in the model.
Although the best $r^2$ scores of the final models reach, on average, 0.7 for the training sets, we only get, on average, a best score of 0.5 for the test sets (Fig. \ref{PredTask}). The method seems to reach a limit. One also observes that even if a model better fits the training set, it does not mean that it will give the best recovery. Indeed, it is sometimes better to consider a model with fewer regressors, and so a lower $r^2$ for training set, to better recover. 

\subsection{Bibliographic data.}
\label{Experiments_Same}

Bibliographic networks are also good examples of heterogeneous information networks since they contain multiple types of nodes and links. We here focus on scientific publications.

\subsubsection{Data Set Description and Setup.}
\begin{figure}[h]
\captionsetup[subfigure]{position=top, labelfont=bf,textfont=normalfont,singlelinecheck=off,justification=raggedright}
\centering
\subfloat[]{\includegraphics[scale=0.5]{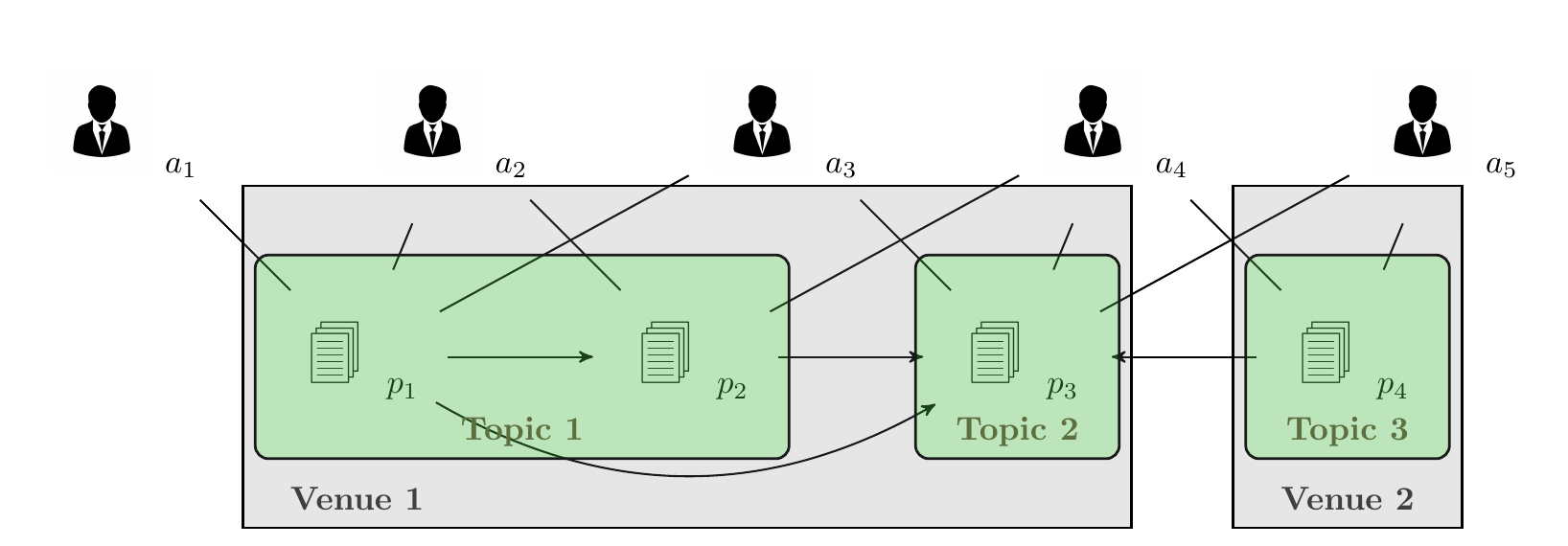} \label{Schema_bibneta}}

\vspace*{0.5cm}

\subfloat[]{\includegraphics[scale=0.7]{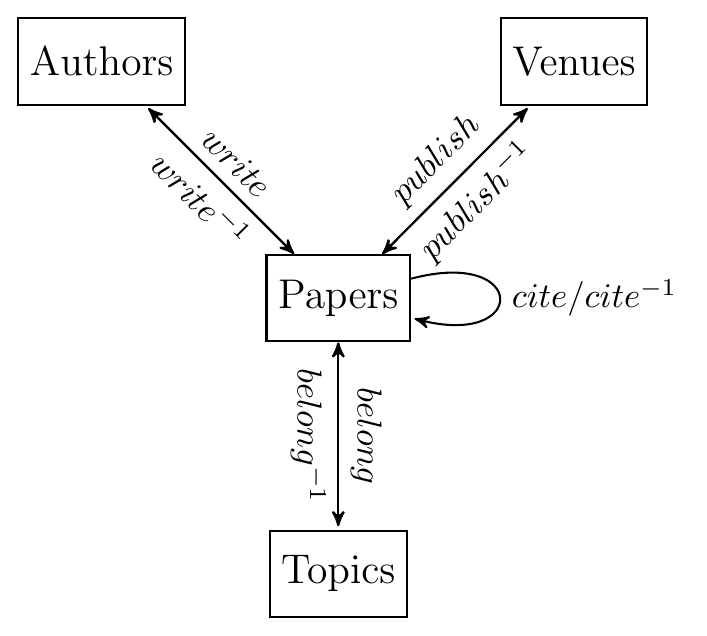} \label{Schema_bibnetb}}
\caption{\textbf{(a)} Example of a bibliographic network. 
\textbf{(b)} Its associated network schema.}
\label{Schema_bibnet}
\end{figure}

Fig. \ref{Schema_bibnet} illustrates an example of such networks where one can distinguish four types of nodes that is authors, papers, venues and topics; and four types of links (eight when we differentiate a type from its inverse) that is write, publish, cite and belong to.

The HIN we analyse in this article is constructed from DBLP publications \cite{aminer}.
The data set contains 95,855 authors with 1,537,407 co-author relationships and 186,175 papers with 1,356,893 citation relationships.
The papers belong to nine distinct topics: Artificial Intelligence, Computer Graphic: multimedia, Computer Networks, Database: Data Mining: Information Retrieval, Human Computer Interaction: Ubiquitous Computing, Information Security, Interdisciplinary Studies, Software Engineering and Theoretical Computer Science. These topics are represented in the 92 venues present in the data set. 

The presented method is used to find out if the co-author relationship A$\rightarrow$P$\leftarrow$A is correlated with other directly extractable relationships of the underlying graph. 
Table \ref{Table_metapaths} shows the different meta-paths used in the models selected according to their semantics contrary to the previous experiment. 
Since there is only one directed type of links between two given types of nodes, we only mention the types of nodes to describe the meta-path. \\

\begin{table}[h!]
\centering
\begin{small}
\renewcommand*{\arraystretch}{0.95}
\begin{tabular}{lll}
\hline
\rowcolor{gray!50}
Meta-Path & Meaning & Feature \\
\hline 
\hline
\textbf{A$\rightarrow$P$\leftarrow$A} & \textbf{are co-authors} & \\

\rowcolor{gray!25}
A$\rightarrow$P$\leftarrow$A$\rightarrow$P$\leftarrow$A & share co-authors\footnotemark & $v_A$ \\

A$\rightarrow$P$\rightarrow$P$\leftarrow$A & cite the other's paper & \multirow{2}{*}{$v_{PP}$} \\
A$\rightarrow$P$\leftarrow$P$\leftarrow$A & are cited by the other's paper & \\

\rowcolor{gray!25}
A$\rightarrow$P$\rightarrow$P$\leftarrow$P$\leftarrow$A & co-cite the same paper & \\
\rowcolor{gray!25}
A$\rightarrow$P$\leftarrow$P$\rightarrow$P$\leftarrow$A & are co-cited by the same paper & \multirow{-2}{*}{$v_{PPP}$} \\

A$\rightarrow$P$\rightarrow$V$\leftarrow$P$\leftarrow$A & have paper in the same conference & $v_V$ \\

\rowcolor{gray!25}
A$\rightarrow$P$\rightarrow$T$\leftarrow$P$\rightarrow$A & have paper about the same topic & $v_T$\\
\hline
\end{tabular}
\end{small}
\caption{Meta-paths describing some notions of proximity between authors. The Features gather some meta-paths that are similar if the direction of the arrows is neglected or alternatively, if one only considers the node types composing the meta-paths.}
\label{Table_metapaths}
\end{table}
\footnotetext{distinct of the targeted authors}

As mentioned, meta-paths are no longer determined by their length but selected by a more solid prior knowledge of the data.
Here are given some motivations about the selected meta-paths.
\begin{itemize}
\setlength\itemsep{-0.2em}
\item A$\rightarrow$P$\rightarrow$A$\leftarrow$P$\rightarrow$A means that two authors have written with a third common author. It represents a triangle when the AP-PA graph is projected onto A. This meta-path is the most ``social'';
\item A$\rightarrow$P$\rightarrow$P$\leftarrow$A and A$\rightarrow$P$\leftarrow$P$\leftarrow$A state for the interest of a person (say $a$) for the work of another (say $b$). It could be meaningful to think that if $a$ is interested in $b$'s work and cites it, $a$ is eager to communicate with $b$ and even to collaborate and to publish with her. The same holds if $a$ and $b$ exchange their role;
\item A$\rightarrow$P$\rightarrow$P$\leftarrow$P$\leftarrow$A means that two authors cite the same paper and are thus inspired by the same ideas. This could be a good reason for a co-author relation;
\item A$\rightarrow$P$\leftarrow$P$\rightarrow$P$\leftarrow$A is quite different since it states that a third person (say $c$) cites the work of $a$ et $b$ but it does not mean that $a$ and $b$ work on the same thing. So, we expect this meta-path to be less significant that the previous one, albeit the structure is fairly close;
\item A$\rightarrow$P$\rightarrow$V$\leftarrow$P$\leftarrow$A and A$\rightarrow$P$\rightarrow$T$\leftarrow$P$\leftarrow$A mean that $a$'s paper and $b$'s paper are in the same venue or belong to the same topic respectively. Even if some venues can gather a lot of people, being accepted in the same venue might trigger collaborations. Plus, working on the same topic can also be a source of collaboration. 
\end{itemize}

Starting from the data, we construct four matrices associated to four bipartite graphs. 
In particular, $AP$ where $AP_{ap}$ equals 1 when authors $a$ writes paper $p$, 0 otherwise. $PP$ where $PP_{pq}$ equals 1 when paper $p$ cites paper $q$, 0 otherwise. $PV$ where $PV_{pv}$ equals 1 when paper $p$ is published/presented in conference/venue $v$, 0 otherwise. $PT$ where $PT_{pt}$ equals 1 when paper $p$ belongs to topic $t$, 0 otherwise. 
These matrices are binary but it does not imply the co-author matrix ($AA$) is binary too.
In order to compute the proposed variables/meta-paths, matrices are transformed into row-stochastic matrices i.e., normalized such that the sum of each line equals 1. In this setting, we can consider these matrices as transition matrices and perform random walks on it. 
For Fig. \ref{Schema_bibneta}, we have the following matrices:
\begin{center}
\begin{footnotesize}
\begin{tabular}{ll}
$
AP = 
\begin{bmatrix}
1 & 0 & 0 & 0 & \textcolor{red}{0}\\
1/2 & 1/2 & 0 & 0 & \textcolor{red}{0}\\
1/2 & 0 & 1/2 & 0 & \textcolor{red}{0}\\
0 & 1/3 & 1/3 & 1/3 & \textcolor{red}{0}\\
0 & 0 & 1/2 & 1/2 & \textcolor{red}{0}\\ 
\textcolor{red}{0} & \textcolor{red}{0} & \textcolor{red}{0} & \textcolor{red}{0} & \textcolor{red}{1} 
\end{bmatrix}
$
&
$
PV = 
\begin{bmatrix}
1 & 0 & \textcolor{red}{0}\\
1 & 0 & \textcolor{red}{0}\\
1 & 0 & \textcolor{red}{0}\\
0 & 1 & \textcolor{red}{0}\\
\textcolor{red}{0} & \textcolor{red}{0} & \textcolor{red}{1}
\end{bmatrix}
$ \\ \\
$
PP = 
\begin{bmatrix}
0 & 1/2 & 1/2 & 0 & \textcolor{red}{0} \\
0 & 0 & 1 & 0 & \textcolor{red}{0} \\
0 & 0 & 0 & 0 & \textcolor{red}{1} \\
0 & 0 & 1 & 0 & \textcolor{red}{0} \\
\textcolor{red}{0} & \textcolor{red}{0} & \textcolor{red}{0} & \textcolor{red}{0} & \textcolor{red}{1}
\end{bmatrix}
$&$
PT = 
\begin{bmatrix}
1 & 0 & 0 & \textcolor{red}{0} \\
1 & 0 & 0 & \textcolor{red}{0} \\
0 & 1 & 0 & \textcolor{red}{0} \\
0 & 0 & 1 & \textcolor{red}{0} \\
\textcolor{red}{0} & \textcolor{red}{0} & \textcolor{red}{0} & \textcolor{red}{1}  
\end{bmatrix}
$
\end{tabular}
\end{footnotesize}
\end{center}

\noindent where the red entries (last columns and rows of each matrix) are related to the so-called hole nodes (see Sec. \ref{Subsection_PCRW}). Remark that paper $p_3$ points to the hole node in the $PP$ graph since it does not cite any paper.

Furthermore, note that for meta-paths of the form A$\rightarrow$P$\rightarrow$ ``node type'' $\leftarrow$P$\leftarrow$A with ``node type'' in $\{$P, V, T$\}$, one forbids the walker to return to the same paper in his second and fourth step. 
It prevents us from using what we are looking for.
For instance, in Fig. \ref{Schema_bibnet}, a walker constrained by the A$\rightarrow$P$\rightarrow$A$\leftarrow$P$\leftarrow$A meta-path and having travelled through the path $a_1 \rightarrow p_1 \rightarrow a_3$ cannot return on $p_1$ at her next step but has to go to $p_3$.

\subsubsection{Results.}

As said, the aim of this experiment is to express the distribution of co-author relationship of all the authors in the data set by a combination of other distributions. The results are once again divided into explanatory and recovery tasks.

\textbf{Forward Liner Regression for Data Description.}
Two tests are performed: first, we consider all the presented meta-paths as regressors (Table \ref{Table_results_test1}) and second, we aggregate some meta-paths a.k.a. features (see third column of Table \ref{Table_metapaths}) and utilize them into the algorithm (Table \ref{Table_results_test2}). 
We propose this aggregation because if the direction of the arrows is neglected, the meta-paths composing a feature are the same. In other words, the sequence of the node types is the same. The aim is to quantify the quality loss (if any) of the prediction when aggregating meta-paths into features.

\textit{Meta-paths as regressors.}
For the first test, we see that only three meta-paths are retained into the final model. This latter is able to explained 66,61\% of the variance
in the dependent variable from the independent variables. According to this model, the most significant meta-paths to explained the co-author relationship are related to the way authors share the same co-authors (some kind of transitivity\footnote{Transitivity of the authors-authors graph equals 0.6948.}), cite and co-cite, plus the venues in which papers are published/presented. Meta-path related to ``topic'' is not included in the model.

\begin{table}[h!]
\centering
\renewcommand*{\arraystretch}{0.95}
\begin{tabular}{lll}
\hline
\rowcolor{gray!50}
Meta-Path & Coefficient & $p$-value \\
\hline 
\hline
A$\rightarrow$P$\leftarrow$A$\rightarrow$P$\leftarrow$A & 1.2507 & 0.0038 \\

\rowcolor{gray!25}
A$\rightarrow$P$\rightarrow$P$\leftarrow$A & 0.9237 & 0.0099 \\
\rowcolor{gray!25}
A$\rightarrow$P$\leftarrow$P$\leftarrow$A & - & -  \\

A$\rightarrow$P$\rightarrow$P$\leftarrow$P$\leftarrow$A & 0.2813 & 0.0395 \\
A$\rightarrow$P$\leftarrow$P$\rightarrow$P$\leftarrow$A & - & - \\

\rowcolor{gray!25}
A$\rightarrow$P$\rightarrow$V$\leftarrow$P$\leftarrow$A & 0.1539 & 0.0099 \\

A$\rightarrow$P$\rightarrow$T$\leftarrow$P$\leftarrow$A & - & - \\
\hline
\rowcolor{gray!35}
$r^2$ & \multicolumn{2}{c}{0.6661} \\
\hline
\end{tabular}
\caption{Results of the linear model for all selected meta-paths.}
\label{Table_results_test1}
\end{table}

\textit{Features as regressors.}
When we aggregated some meta-paths into features, those related to citing the same paper and the venues are not included in the model (see Table \ref{Table_results_test2}). For the first one, it could be explained by the fact that only one meta-path (A$\rightarrow$P$\rightarrow$P$\leftarrow$P$\leftarrow$A) among two is imported in the first test\footnote{Of course, the same remark can be made for the $v_{PP}$ meta-paths and yet, $v_{PP}$ is part of the model.}. 
\begin{table}[!b]
\centering
\renewcommand*{\arraystretch}{0.95}
\begin{tabular}{lll}
\hline
\rowcolor{gray!50}
Feature & Coefficient & $p$-value \\
\hline 
\hline
$v_A$ & 1.2133 & 0.0028 \\

\rowcolor{gray!25}
& & \\
\rowcolor{gray!25}
\multirow{-2}{*}{$v_{PP}$} & \multirow{-2}{*}{1.8549} & \multirow{-2}{*}{0.0034} \\

\multirow{2}{*}{$v_{PPP}$} & \multirow{2}{*}{-} & \multirow{2}{*}{-} \\
& & \\

\rowcolor{gray!25}
$v_V$ & - & - \\

$v_T$ & - & - \\
\hline
\rowcolor{gray!35}
$r^2$ & \multicolumn{2}{c}{0.5997} \\
\hline
\end{tabular}
\caption{Results of the linear model for meta-paths aggregated into features.}
\label{Table_results_test2}
\end{table}
No immediate reason is given for the absence of $v_V$ variable. 
Plus, this second model only accounts for 59.97\% of the variance: each meta-path brings its own meaning and even if some of them seem close to each other, wanting to aggregate them is not beneficial for our purpose. Actually, we have already mentioned a ``fundamental'' difference between variables of $v_{PPP}$.
As in the previous case, feature related to ``topic'' is not significant for the specific objective when other variables (see Table \ref{Table_metapaths}) are considered in the forward linear regression.

\textit{Topics under scrutiny.}
The small number of considered topics, compared with the number of papers, could partly explain why the topic meta-path is not taken into account. Indeed, only one topic is assigned to each paper so the meta-path P$\rightarrow$T$\rightarrow$P generate a dense ``paper-paper matrix''\footnote{The same comment could be made for meta-path P$\rightarrow$V$\rightarrow$P since the number of venues is also limited - although to a lesser extent since a topic encompasses several venues. The number of non zero entries of the matrix APTPA (not really the same as PTP but the final result is encompassed in APTPA) equals 6,515,232 while for APVPA, this number raises to 3,940,634, which is still 1.6 times lower.} and when computing the matrix product  
A$\rightarrow$P$\rightarrow$T$\leftarrow$P$\leftarrow$A, any relevant information is somewhat lost.

Thus we think the meta-path A$\rightarrow$P$\rightarrow$T$\leftarrow$P$\leftarrow$A brings a too diffuse information. However, the idea of considering topics is not meaningless since an author interested in a topic is often interested for a while and therefore, has the time to collaborate with other people, who are themselves interested in the same subject. Authors writing about a same topic might partly be co-authors.  

\begin{table}[h]
\centering
\renewcommand*{\arraystretch}{0.95}
\rowcolors{2}{gray!25}{white}
\begin{tabular}{lcccc}
\hline
\rowcolor{gray!50}
Topic & \#auth. & \#pap. & \#ven. & $r^2$ \\
\hline
\hline
A. I. & 41538 & 65927 & 23 & 0.5914 \\ 
Comp. Graph. Mult. & 25989 & 18877 & 13 & 0.6358 \\
Comp. Net. & 22374 & 30212 & 9 & 0.6321 \\
Database & 5865 & 9294 & 7 & 0.7349 \\
Hum. Comp. Inter. & 4660 & 10666 & 5 & 0.7723 \\
Info. Sec. & 5298 & 6943 & 6 & 0.7211 \\
Interdisc. Std. & 46111 & 2614 & 11 & 0.7838 \\
Software Eng. & 8147 & 20506 & 8 & 0.7222 \\
Th. Comp. Sci. & 10824 & 21136 & 11 & 0.5796 \\
\hline
\end{tabular}
\caption{Results of the different topics.}
\label{Fig_res_sep}
\end{table}

\begin{figure}[!b]
\centering
\includegraphics[scale=0.34]{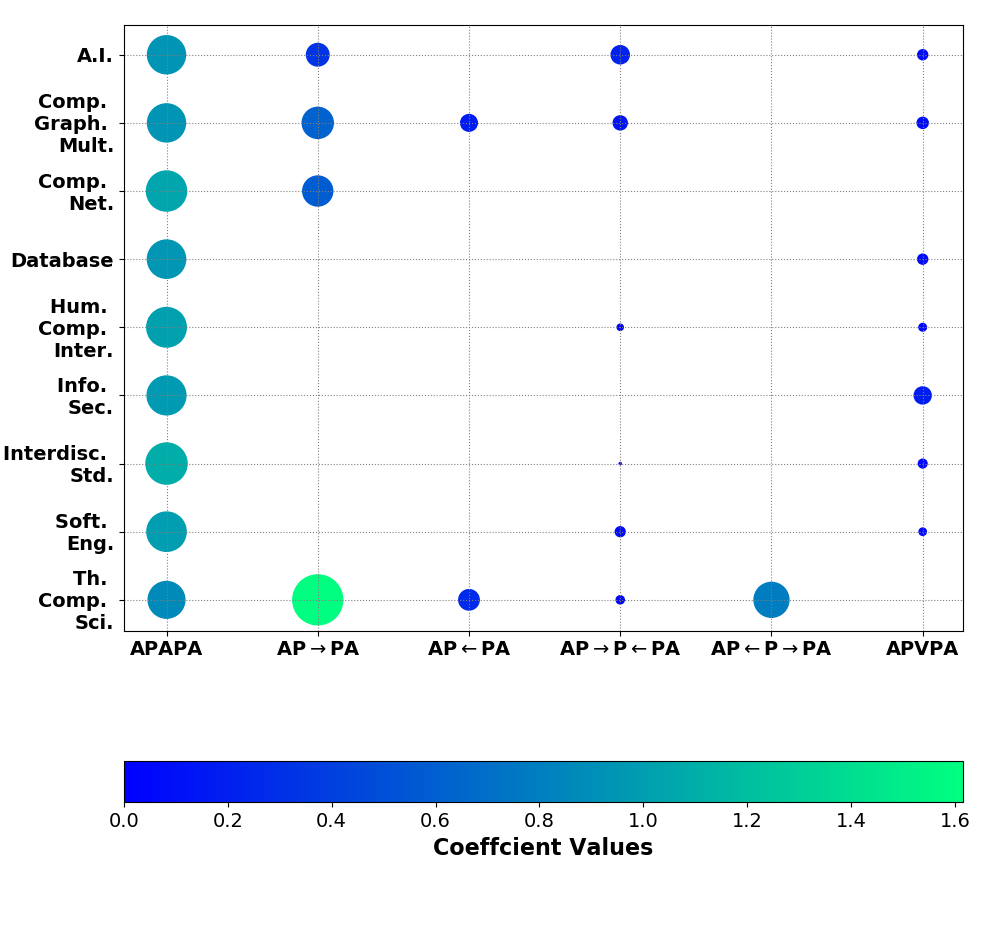}
\caption{Coefficient values of the final models of the different domains. Size of the bullets is proportional to the coefficient values.}
\label{Multi_Bar}
\end{figure}

So, we split the data into nine subsets, each one related to one topic, and apply the method with the six meta-paths cited above i.e., all except A$\rightarrow$P$\rightarrow$T$\leftarrow$P$\leftarrow$A (see Remark \ref{Division_NT}). Results are reported in Table \ref{Fig_res_sep}. 
On average, we have a better descriptive model than before: $\langle r^2 \rangle = 0.6970$ (and $\sigma$= 0.0710). 
This could mean that inside some topics, there are some patterns more homogeneous or frequent and we are more capable of explaining them.
However, for Artificial Intelligence and Theoretical Computer Science, it is harder to find a model that fits the data well.

In Fig. \ref{Multi_Bar} are reported the final models' coefficient values for the different topics.
Meta-path A$\rightarrow$P$\rightarrow$A$\leftarrow$P$\leftarrow$A is selected by each topic: sharing the same co-author is the most useful to explain co-author relationship in a given topic. 
Meta-path A$\rightarrow$P$\rightarrow$V$\leftarrow$P$\leftarrow$A is important for 7 topics out of 9. Only Computer Networks and Theoretical Computer Science do not take it into account. 
Note that only Theoretical Computer Science includes A$\rightarrow$P$\leftarrow$P$\rightarrow$P$\leftarrow$A in its final model. This topic is also the only one for which A$\rightarrow$P$\rightarrow$A$\leftarrow$P$\leftarrow$A has not the greatest coefficient, it is surpassed by A$\rightarrow$P$\rightarrow$P$\leftarrow$A and closely followed by A$\rightarrow$P$\rightarrow$P$\leftarrow$P$\leftarrow$A. The paper relations seem highly important for this domain. 

\textbf{Forward Linear Regression for Data Recovery.}
We are now interested in the recovery of link weights. Average results Monte-Carlo cross-validations are reported in Table \ref{RecoveryDomains}. All $p$-values associated to the regressors are below the fixed threshold $\alpha=0.05$.

\begin{table}[!b]

\centering
\begin{tabular}{llc}
\hline
\rowcolor{gray!50}
Topics & Meta-Paths & $\langle r^2_{test} \rangle$\\
\hline
\hline
& A$\rightarrow$P$\leftarrow$A$\rightarrow$P$\leftarrow$A & \\
& A$\rightarrow$P$\rightarrow$P$\leftarrow$A & \\
\multirow{-3}{*}{All topics} & A$\rightarrow$P$\leftarrow$V$\rightarrow$P$\leftarrow$A  & \multirow{-3}{*}{0.5508} \\ 

\hline
\rowcolor{gray!25}
& A$\rightarrow$P$\leftarrow$A$\rightarrow$P$\leftarrow$A & \\
\rowcolor{gray!25}
\multirow{-2}{*}{A.I} & A$\rightarrow$P$\rightarrow$P$\leftarrow$P$\leftarrow$A & \multirow{-2}{*}{0.4994} \\

& A$\rightarrow$P$\leftarrow$A$\rightarrow$P$\leftarrow$A & \\
& A$\rightarrow$P$\rightarrow$P$\leftarrow$A & \\
\multirow{-3}{*}{Comp. Graph. Mult.} & A$\rightarrow$P$\rightarrow$P$\leftarrow$P$\leftarrow$A & \multirow{-3}{*}{0.5133} \\

\rowcolor{gray!25}
& A$\rightarrow$P$\leftarrow$A$\rightarrow$P$\leftarrow$A & \\ 
\rowcolor{gray!25}
\multirow{-2}{*}{Comp. Net.} & A$\rightarrow$P$\rightarrow$P$\leftarrow$A & \multirow{-2}{*}{0.5322} \\

& A$\rightarrow$P$\leftarrow$A$\rightarrow$P$\leftarrow$A & \\
\multirow{-2}{*}{Database} & A$\rightarrow$P$\leftarrow$V$\rightarrow$P$\leftarrow$A & \multirow{-2}{*}{0.7258} \\

\rowcolor{gray!25}
& A$\rightarrow$P$\leftarrow$A$\rightarrow$P$\leftarrow$A & \\
\rowcolor{gray!25}
& A$\rightarrow$P$\leftarrow$P$\rightarrow$P$\leftarrow$A & \\
\rowcolor{gray!25}
\multirow{-3}{*}{Hum. Comp. Inter.} & A$\rightarrow$P$\rightarrow$V$\leftarrow$P$\leftarrow$A & \multirow{-3}{*}{0.7338} \\

& A$\rightarrow$P$\leftarrow$A$\rightarrow$P$\leftarrow$A & \\
\multirow{-2}{*}{Info. Sec.} & A$\rightarrow$P$\leftarrow$V$\rightarrow$P$\leftarrow$A & \multirow{-2}{*}{0.6509} \\

\rowcolor{gray!25}
& A$\rightarrow$P$\leftarrow$A$\rightarrow$P$\leftarrow$A & \\
\rowcolor{gray!25}
\multirow{-2}{*}{Interdisc. Std.} & A$\rightarrow$P$\leftarrow$V$\rightarrow$P$\leftarrow$A & \multirow{-2}{*}{0.7440} \\

& A$\rightarrow$P$\leftarrow$A$\rightarrow$P$\leftarrow$A & \\
& A$\rightarrow$P$\rightarrow$P$\leftarrow$P$\leftarrow$A & \\
\multirow{-3}{*}{Software Eng.} & A$\rightarrow$P$\leftarrow$V$\rightarrow$P$\leftarrow$A & \multirow{-3}{*}{0.6450} \\

\rowcolor{gray!25}
& A$\rightarrow$P$\leftarrow$A$\rightarrow$P$\leftarrow$A & \\
\rowcolor{gray!25}
& A$\rightarrow$P$\rightarrow$P$\leftarrow$A & \\
\rowcolor{gray!25}
& A$\rightarrow$P$\leftarrow$P$\leftarrow$A & \\
\rowcolor{gray!25}
& A$\rightarrow$P$\rightarrow$P$\leftarrow$P$\leftarrow$A & \\
\rowcolor{gray!25}
\multirow{-5}{*}{Th. Comp. Sci.} & A$\rightarrow$P$\leftarrow$P$\rightarrow$P$\leftarrow$A & \multirow{-5}{*}{0.3557} \\
\hline
\end{tabular}

\caption{Results of the recovery task for the general case (all topics) and per topic.}
\label{RecoveryDomains}
\end{table}

For Database, Human Computer Interaction and Interdisciplinary Studies, the recovery is somehow achievable in the sense that the score of the test set is almost as good as for the training set. For the other domains, the quality loss is more significant, even for Information Security and Software Engineering which have a good $r^2$ for the training set. 
Finally, note that for Theoretical Computer Sciences, the $r^2$ for recovery is really low and its true relevance can be somewhat even questioned (albeit the $p$-values are below 0.05). 
However, to be sure of its relevance, the results computed from our data set are compared with a null hypothesis model that preserve some properties of the network topology (e.g. degree distributions) but randomly reshuffles the links among the nodes. 
The aim is to show that degree distributions only are not enough to generate such a correlation in the data and that this correlation arises from the particular data or at least, from more involved topological properties. Indeed, results for such null models are not significant (no regressor with $p$-value smaller than 0.12) and the average score $\langle r^2 \rangle$ over 15 generations of null graphs are at most equal to 0.26.

\section{RELATED WORK}
\label{RelWork}

Compared to previous work, which usually focuses on undirected binary 
graphs, the approach we present addresses the recovery of 
\emph{directed} and \emph{weighted} links in HIN. To this end, our 
regression model directly estimates the weight of links without 
computing any intermediate ranking on these links, or applying any 
threshold to reduce the recovery task to binary graphs.

As previously explained, our work is based on node similarity measures and thus, is also related to link prediction. 
Similarity measures and link prediction have been extensively studied in the past few years.
One often roughly differentiates two kinds of approaches: unsupervised versus supervised.
For the first category, one often proposes different similarity measures based upon either node attributes or the topology of the underlying graph. One can further distinguish local from global indices. Local indices makes use of local neighborhood information e.g. Adamic-Adar index, Common Neighbor or Preferential Attachment Index, Ressource Allocation just to name a few. By contrast, global indices are based on global properties such as paths. These encompass Shortest Path, Katz or measures using random walks e.g. Random Walk with Restart, PageRank, Hitting Time, Commute Time and so on.   
Based on these aforementioned features, a plethora of supervised methods have been conceived to predict links. Amongst them, one distinguishes feature-based classification \cite{al2006link, raymond2010fast} from probabilistic model \cite{leroy2010cold, wang2007local} and matrix factorization \cite{menon2011link}. 
However, all these measures are mostly used in homogeneous networks and for a review of these methods, see \cite{lu2011link, martinez2017survey}.

Recently, several measures have tackled the problem of node similarity in HIN which takes into account not only the structure similarity of two entities but also the metapaths connecting them. Amongst these measures, PathCount (PC\cite{sun2011co}) and Path Constrained Random Walk (PCRW\cite{lao2010relational}) are the two most basic and gave birth to several extensions \cite{fang2016semantic, gupta2017dprel, huang2016meta, zhou2017recurrent}.

Methods related to PC are based on the count of paths between a pair of nodes, given a meta path. PathSim \cite{sun2011pathsim} measures the similarity between two objects of same type along a symmetric meta path which is restrictive since many valuable paths are asymmetric and
the relatedness between entities of different types is not useless. Two measures based on it \cite{he2014exploiting,yao2014pathsimext} incorporate more information such as the node degree and the transitivity. However, all these methods have the drawback of favoring highly connected objects since they deal with raw data.
 
Methods related to PCRW are based on random walks and so the probability of reaching a node from another one, given a meta path. Considering a random walk implies a normalisation and, depending on the data, offers better results. 
An adaptation, HeteSim \cite{shi2014hetesim}, measures the meeting probability between two walkers starting from opposite extremities of a path, given a meta path. However, this method requires the decomposition of atomic relations for odd-length meta-paths. This decomposition allows the walkers to meet at the middle of the meta-path and at the same node type but it is very costly for large graphs. 
To address this issue, AvgSim \cite{meng2014relevance} computes the similarity between two nodes using random walks conditioned by a meta path and its inverse. But it is mostly appreciated in undirected networks since in these cases, it is just as sensible to walk a path in one direction as in the other.

In these cited works, when the similarity scores are used for link prediction/detection, the scores are ranked and then, the presence of links is inferred based on this ranking. 
Also some work try to combine meta-paths but the target values to recover are binary; the networks are unweighted. 
At variance with these works, we set ourselves in the general framework of directed and weighted HINs. We do not use any ranking or threshold but take directly the similarity measures obtained by means of an adequate combination of PCRWs as link weights. This allows not only to perform description tasks but also, to some extent, recovery tasks.

\section{CONCLUSION}
\label{Conclusion}

We have considered a linear combination of probability distributions resulting from path-constrained random walks to explain, to some extent, a specific relation in heterogeneous information networks. 
This proposed method allows to express the weight of a link between two nodes knowing some other links in a graph. 
This could be useful for prediction or recommendation tasks.
In particular, we have shown by working on Twitter data, that the hashtags posted by a specific user is mainly related to those posted by her direct neighborhood, especially the mention and reply neighborhood. This method has also shown that the retweet relation is not really useful for our purpose.
Then we have shown the applicability of the method to bibliographic data in order to recover the co-author relationship. It has been found that (data separated into) some topics are more suited to our method and so, the functioning of co-authors seemed to differ from one topic to another.  

Nevertheless, the main drawback of the method is its sensitivity to outliers. Hence, more robust least square alternatives could be envisaged such that Least Trimmed Squares or parametric alternatives.

Furthermore when there is no prior knowledge about the data, as for the Twitter data experiment, we had to provide all the meta-paths whose length is no longer than four. Even if it has been motivated by previous tests, this threshold is clearly data related. Hence, it could be interesting to build a method able to find relevant meta-paths by itself.

Finally, all data have been aggregated in time. Consequently, the chronology of the events is ignored. Since it is possible to extract the time stamp of tweets or to take into account the papers' publication date, a future work could be the integration of time by defining a random walk process on temporal graphs \cite{masuda2017random} or by counting the temporal paths \cite{latapy2018stream} (plus normalisation). The walker can thus only follow time-respecting paths which can perhaps improve the quality of the model.

\section*{ACKNOWLEDGEMENTS}

This work is funded in part by the European Commission H2020 FETPROACT 
2016-2017 program under grant 732942 (ODYCCEUS)

\bibliographystyle{plain}
\bibliography{Ref_Article}

\begin{thebibliography}{10}

\bibitem{aminer}
\url{https://aminer.org/citation}.

\bibitem{al2006link}
Mohammad Al~Hasan, Vineet Chaoji, Saeed Salem, and Mohammed Zaki.
\newblock Link prediction using supervised learning.
\newblock In {\em SDM06: workshop on link analysis, counter-terrorism and
  security}, 2006.

\bibitem{arlot2010survey}
Sylvain Arlot, Alain Celisse, et~al.
\newblock A survey of cross-validation procedures for model selection.
\newblock {\em Statistics surveys}, 4:40--79, 2010.

\bibitem{fang2016semantic}
Yuan Fang, Wenqing Lin, Vincent~W Zheng, Min Wu, Kevin Chen-Chuan Chang, and
  Xiao-Li Li.
\newblock Semantic proximity search on graphs with metagraph-based learning.
\newblock In {\em 2016 IEEE 32nd International Conference on Data Engineering
  (ICDE)}, pages 277--288. IEEE, 2016.

\bibitem{gupta2017dprel}
Mukul Gupta, Pradeep Kumar, and Bharat Bhasker.
\newblock Dprel: a meta-path based relevance measure for mining heterogeneous
  networks.
\newblock {\em Information Systems Frontiers}, pages 1--17, 2017.

\bibitem{he2014exploiting}
Jiazhen He, James Bailey, and Rui Zhang.
\newblock Exploiting transitive similarity and temporal dynamics for similarity
  search in heterogeneous information networks.
\newblock In {\em International Conference on Database Systems for Advanced
  Applications}, pages 141--155. Springer, 2014.

\bibitem{huang2016meta}
Zhipeng Huang, Yudian Zheng, Reynold Cheng, Yizhou Sun, Nikos Mamoulis, and
  Xiang Li.
\newblock Meta structure: Computing relevance in large heterogeneous
  information networks.
\newblock In {\em Proceedings of the 22nd ACM SIGKDD International Conference
  on Knowledge Discovery and Data Mining}, pages 1595--1604. ACM, 2016.

\bibitem{lao2010relational}
Ni~Lao and William~W Cohen.
\newblock Relational retrieval using a combination of path-constrained random
  walks.
\newblock {\em Machine learning}, 81(1):53--67, 2010.

\bibitem{latapy2018stream}
Matthieu Latapy, Tiphaine Viard, and Cl{\'e}mence Magnien.
\newblock Stream graphs and link streams for the modeling of interactions over
  time.
\newblock {\em Social Network Analysis and Mining}, 8(1):61, 2018.

\bibitem{leroy2010cold}
Vincent Leroy, B~Barla Cambazoglu, and Francesco Bonchi.
\newblock Cold start link prediction.
\newblock In {\em Proceedings of the 16th ACM SIGKDD international conference
  on Knowledge discovery and data mining}, pages 393--402. ACM, 2010.

\bibitem{lu2011link}
Linyuan L{\"u} and Tao Zhou.
\newblock Link prediction in complex networks: A survey.
\newblock {\em Physica A: statistical mechanics and its applications},
  390(6):1150--1170, 2011.

\bibitem{macskassy2012study}
Sofus~A Macskassy.
\newblock On the study of social interactions in twitter.
\newblock In {\em Sixth International AAAI Conference on Weblogs and Social
  Media}, 2012.

\bibitem{martinez2017survey}
V{\'\i}ctor Mart{\'\i}nez, Fernando Berzal, and Juan-Carlos Cubero.
\newblock A survey of link prediction in complex networks.
\newblock {\em ACM Computing Surveys (CSUR)}, 49(4):69, 2017.

\bibitem{masuda2017random}
Naoki Masuda, Mason~A Porter, and Renaud Lambiotte.
\newblock Random walks and diffusion on networks.
\newblock {\em Physics reports}, 716:1--58, 2017.

\bibitem{meng2014relevance}
Xiaofeng Meng, Chuan Shi, Yitong Li, Lei Zhang, and Bin Wu.
\newblock Relevance measure in large-scale heterogeneous networks.
\newblock In {\em Asia-Pacific Web Conference}, pages 636--643. Springer, 2014.

\bibitem{menon2011link}
Aditya~Krishna Menon and Charles Elkan.
\newblock Link prediction via matrix factorization.
\newblock In {\em Joint european conference on machine learning and knowledge
  discovery in databases}, pages 437--452. Springer, 2011.

\bibitem{raymond2010fast}
Rudy Raymond and Hisashi Kashima.
\newblock Fast and scalable algorithms for semi-supervised link prediction on
  static and dynamic graphs.
\newblock In {\em Joint european conference on machine learning and knowledge
  discovery in databases}, pages 131--147. Springer, 2010.

\bibitem{shi2014hetesim}
Chuan Shi, Xiangnan Kong, Yue Huang, S~Yu Philip, and Bin Wu.
\newblock Hetesim: A general framework for relevance measure in heterogeneous
  networks.
\newblock {\em IEEE Transactions on Knowledge and Data Engineering},
  26(10):2479--2492, 2014.

\bibitem{shi2016survey}
Chuan Shi, Yitong Li, Jiawei Zhang, Yizhou Sun, and S~Yu Philip.
\newblock A survey of heterogeneous information network analysis.
\newblock {\em IEEE Transactions on Knowledge and Data Engineering},
  29(1):17--37, 2016.

\bibitem{sun2011co}
Yizhou Sun, Rick Barber, Manish Gupta, Charu~C Aggarwal, and Jiawei Han.
\newblock Co-author relationship prediction in heterogeneous bibliographic
  networks.
\newblock In {\em 2011 International Conference on Advances in Social Networks
  Analysis and Mining}, pages 121--128. IEEE, 2011.

\bibitem{sun2011pathsim}
Yizhou Sun, Jiawei Han, Xifeng Yan, Philip~S Yu, and Tianyi Wu.
\newblock Pathsim: Meta path-based top-k similarity search in heterogeneous
  information networks.
\newblock {\em Proceedings of the VLDB Endowment}, 4(11):992--1003, 2011.

\bibitem{wang2007local}
Chao Wang, Venu Satuluri, and Srinivasan Parthasarathy.
\newblock Local probabilistic models for link prediction.
\newblock In {\em Seventh IEEE international conference on data mining (ICDM
  2007)}, pages 322--331. IEEE, 2007.

\bibitem{xu2001monte}
Qing-Song Xu and Yi-Zeng Liang.
\newblock Monte carlo cross validation.
\newblock {\em Chemometrics and Intelligent Laboratory Systems}, 56(1):1--11,
  2001.

\bibitem{yao2014pathsimext}
Kun Yao, Hoi~Fong Mak, et~al.
\newblock Pathsimext: revisiting pathsim in heterogeneous information networks.
\newblock In {\em International Conference on Web-Age Information Management},
  pages 38--42. Springer, 2014.

\bibitem{zhou2017recurrent}
Yu~Zhou, Jianbin Huang, Heli Sun, and Yizhou Sun.
\newblock Recurrent meta-structure for robust similarity measure in
  heterogeneous information networks.
\newblock {\em arXiv preprint}, 2017.

\end{thebibliography}

\end{document}